\documentclass[preprint]{revtex4-2}
\usepackage{xcolor}
\usepackage{graphicx}% Include figure files
\usepackage{bm}% bold math

\usepackage{filecontents}
\usepackage{siunitx}
\DeclareSIUnit\torr{Torr}
\DeclareSIUnit\sccm{sccm}
\usepackage{mhchem}
\usepackage[utf8]{inputenc}
\usepackage[T1]{fontenc}
\usepackage{hyperref}
\usepackage{etoolbox}
\usepackage{lineno}

\begin{document}

\title{Lateral Cavity-Enhanced Guided Mode Resonance Structures for Mid-wave Infrared Photodetector Pixels}

\author{Sreeja Purkait}
\altaffiliation{The authors contributed equally to this work}
\affiliation{Department of Physics and Applied Physics\\
  University of Massachusetts Lowell\\
  Lowell, MA 01854}
\author{Noah Mansfield}
\altaffiliation{The authors contributed equally to this work}
\author{Yadviga Tischenko}
\author{Amogh Raju}
\affiliation{Department of Electrical and Computer Engineering\\
  University of Texas Austin\\
  USA}
\author{William Reggio}
\affiliation{Department of Physics and Applied Physics\\
  University of Massachusetts Lowell\\
  Lowell, MA 01854}
\author{Daniel Wasserman}
\affiliation{Department of Electrical and Computer Engineering\\
  University of Texas Austin\\
  USA}
\author{Viktor A. Podolskiy}
\email{viktor_podolskiy@uml.edu}
\affiliation{Department of Physics and Applied Physics\\
  University of Massachusetts Lowell\\
  Lowell, MA 01854}

\begin{abstract} 
We present the design, fabrication, and optical characterization of ultra-compact mid-wave infrared photodetector pixels. Our design relies on a guided mode resonance structure to confine incident mid-infrared light to the \SI{250}{\nm}-thick absorber region of all-epitaxially-grown material stack, and a hybrid cavity-guided mode resonance to confine the mode in the lateral direction. The resulting pixel, with deep subwavelength thickness and lateral dimensions on the order of $\sim 2\lambda_o$ is predicted to achieve external quantum efficiency of the order of 50\%. Our work opens the door for truly compact mid-wave infrared pixels that offer the combined benefits of low dark current, room temperature operation, and small lateral size. 

\end{abstract}

\maketitle
%%%%%%%%%%%%%%%%%%%%%%%%%%  body  %%%%%%%%%%%%%%%%%%%%%%%%%%

\section{Introduction}
The mid-infrared (mid-IR) detector of choice has long leveraged the HgCdTe (MCT) material system, due to its strong absorption, low dark current, and wavelength flexibility \cite{rogalski2005hgcdte}.  However, there has been significant interest, and effort invested, in the development of mid-IR detectors utilizing the more established fabrication infrastructure, more uniform epitaxial growth, and reduced constituent toxicity of III-V semiconductors.  Among the III-V's, type-II superlattice (T2SL) materials\cite{sai1977new,smith1987proposal,osbourn1982strained} offer broad wavelength flexibility, now well-established epitaxial growth, and the potential for incorporation into a range of photodiode and bariode architectures \cite{nBnWicksMaimon, weiss2012inassb,ting2009high}.  While T2SLs have been shown to exhibit decreased Auger scattering \cite{grein1995long,youngdale1994auger,grein2002auger}, and when incorporated into bariode architectures, reduced Shockley-Read-Hall and surface leakage currents \cite{kim2008mid,kim2012long}, they do not possess the large absorption coefficients of bulk MCT \cite{VurgaftmanAPL2016,ROGALSKI2019100228}, and thus the reduction in dark current associated with T2SL-based mid-IR detectors often comes at the expense of reduced responsivity.  Even where bulk materials can be utilized, such as InAsSb lattice-matched to GaSb for the mid-wave infrared (MWIR, $\lambda=3-\SI{5}{\um}$), large dark currents at elevated operating temperatures (i.e. room temperature) limit the competitiveness of mid-IR detectors using III-V semiconductors. 

For this reason, there has been great interest in the design and demonstration of mid-IR detector architectures comprising resonant photonic structures capable of strongly localizing light in ultra-thin device absorbers.  Such systems benefit from greatly reduced dark current, which for diffusion-limited detector operation scales with absorber volume. Simultaneously, as a result of the strongly enhanced absorption resulting from the resonant photonic architecture, these devices do not come with a dramatic responsivity penalty. Recent efforts looking to enhance detector response in ultra-thin absorber volumes have utilized distributed Bragg reflector cavities \cite{RCELetka,RCECanedy}, surface modes on patterned metal gratings\cite{Meyergrating}, and highly-doped semiconductor-based structures supporting surface plasmon polaritons and/or leaky cavity modes\cite{LelandACS,PIQUE,HOT_PIQUE,PPleakycav}.  An alternative approach looks to integrate mid-IR detector architectures into so-called guided-mode resonance (GMR) structures\cite{AvrutskyReflection,WangMagnussonGMR}, where a high-index waveguide core embedding the semiconductor-based detector device, is sandwiched between a low-index bottom cladding layer (leveraging the plasmon response of  highly doped semiconductors) and either air or conductive cladding above the waveguide. Light is coupled into the waveguide mode via a diffraction grating patterned above the detector core, resulting in resonant field confinement at wavelengths determined by the geometry of the GMR and the polarization and angle of the incident light.  By coupling into highly confined laterally propagating modes, a strong enhancement of absorption can be achieved by effectively increasing the interaction length of incident light with the ultra-thin absorber region. These MWIR GMR detectors demonstrate a monolithic architecture capable of strongly enhancing absorption in ultra-thin ($t=\SI{250}{\nano\meter}$) absorber layers, resulting in room temperature device operation with peak MWIR ($\lambda_{0}=\SI{4.1}{\um}$) specific detectivity $D^*>10^{10}$\cite{AKGMR1,AKGMR_2, NCMGMR}.

The driving application for most mid-IR detector development is the demonstration of highly efficient focal plane array (FPA) detectors for a variety of thermal imaging and remote sensing applications \cite{weng2009thermal,neinavaz2021thermal,anding1970estimation}.  Thus, any resonant enhancement mechanism employed must be suitable for FPA implementation and, perhaps more importantly, compatible with IR FPA read-out integrated circuitry (ROICs).  Meeting these conditions requires, first, substrate-side illumination, and second, finite pixel geometries. While the mid-IR detectors demonstrating enhanced response leveraging vertical cavity architectures can translate to smaller lateral pitch, resonant architectures coupling into lateral guided/propagating modes may not fare quite so well upon reduction in lateral size.    
Here we analyze the performance of finite-sized GMR detectors and introduce novel hybrid cavity-GMR geometries that can be used to realize truly compact (few-wavelength-wide and deep-subwavelength-thickness) mid-IR detectors.

\section{Anatomy of Pixels}

\subsection{Detector stack and materials response}
\begin{figure}[t!]
    \centering
    \includegraphics[width=\textwidth]{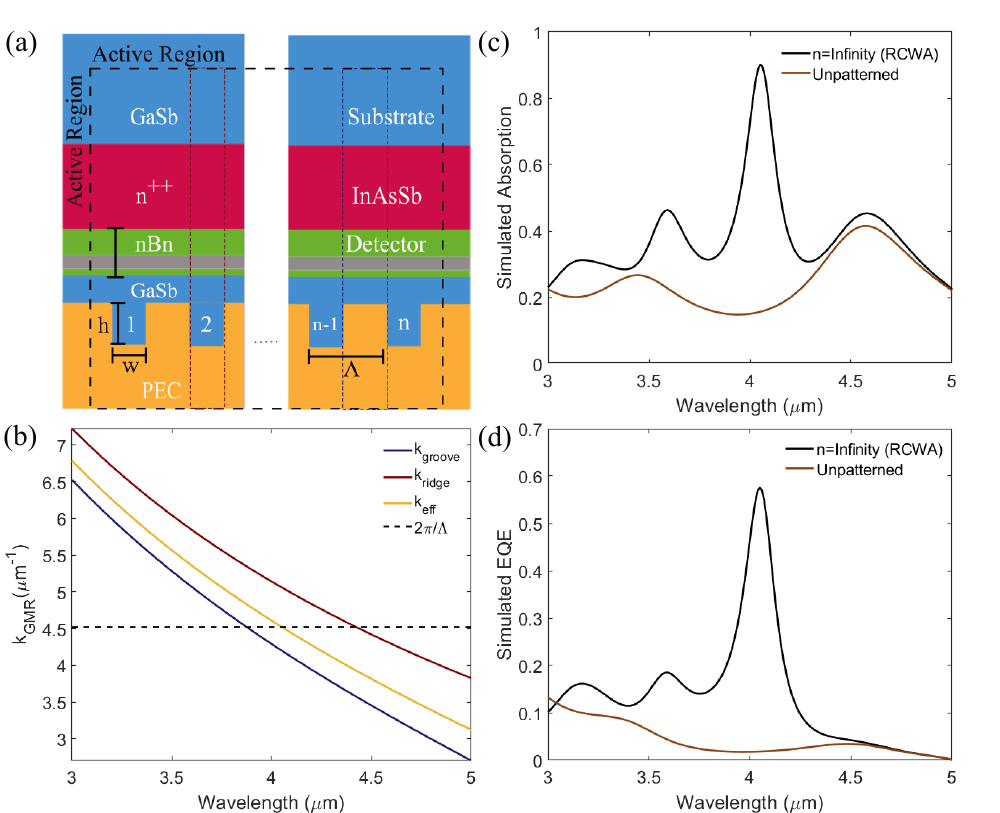}
    \caption{(a) Schematic of the simulated device, with the dashed box indicating the active region used to normalize incident energy flux. (b) Propagation constants of the modes propagating along the unetched (ridge, red) and etched (groove, blue) parts of the device (see dotted regions in  Fig.1(a)) as a function of free space wavelength. Effective propagation constant of the GMR mode (the weighted average of $k_{\rm groove}$ and $k_{\rm ridge}$) is shown with yellow line. The black dashed line indicates reciprocal period of the grating.  (c) Simulated absorption (normalized to incident energy flux) inside the active region for an infinite (n=$\infty$) GMR detector and the planar stack with unpatterned  $\SI{600}{\nano\meter}$-thick GaSb layer. (d)  Simulated EQE  for infinite (n=$\infty$) GMR detector (black) and the planar, unpatterned sample (brown).}
    \label{fig:1}

\end{figure}

The GMR detectors considered in this work utilize the same design as the large area detectors demonstrated in Refs. \citenum{AKGMR1} and \citenum{AKGMR_2}. The detector stack, designed for substrate-side illumination at operating wavelength of $\lambda_o\sim 4 \mu m$ and illustrated in Fig.\ref{fig:1}(a), was grown in a Varian Gen II molecular beam epitaxy (MBE) system on a 2-inch diameter lightly n-doped GaSb substrate, beginning with a $\SI{200}{\nano\meter}$ n-doped GaSb buffer. Following the GaSb buffer layer, a $\SI{1}{\um}$-thick, highly doped ($n^{++}$) T2SL ground plane is grown. This layer plays the role of the low-index cladding for the guided mode used in the detector. The T2SL nBn detector structure,  composed of a $\SI{250}{\nano\meter}$ unintentionally doped (UID) T2SL absorber, a UID AlAs$_{0.08}$Sb$_{0.92}$ $\SI{80}{\nano\meter}$ barrier, and a $\SI{20}{\nano\meter}$ p-type ($N_D=\SI{e18}{cm^{-3}}$) T2SL contact layer, is grown above the n\textsuperscript{++} ground plane. (All T2SL layers have the same composition (InAs/InAs$_{0.49}$Sb$_{0.51}$, with 12 ML period), designed for a cut-off wavelength $\lambda_{\rm cut-off}\simeq {5}{\mu m}$.) Finally, an undoped $\SI{600}{\nano\meter}$ thick GaSb layer is grown above the nBn detector structure, and etched to be used as a grating coupling layer. The detector structure is covered with highly reflective metal (the mid-infrared response of most noble metals is well-approximated by perfect electric conductor boundary conditions) that can be used as one of the electrical contacts in the final device. We assume that the final device will include etching to gain electrical access to the n\textsuperscript{++} layer (see below). In the remainder of the work we focus on the optical absorption in the active layer of the structure that, due to the small thickness of the absorber layer, closely tracks external quantum efficiency (EQE) of the final device\cite{AKGMR1}. With the exception of Sec.3, we do not include the reflection at the substrate/air interface in this analysis, assuming that the final device incorporates an anti-refection coating. 

In our theoretical analysis we note that the extreme conduction band state-filling shifts the effective absorption bandgap of the heavily doped T2SL groundplane into the near-IR\cite{Law:12}. As a result, the optical properties of the n\textsuperscript{++} layer are well described across the mid-IR by the Drude model,    
\begin{equation}
 \epsilon(\omega)=\epsilon_b\left(1-\frac{\omega_p^2}{\omega^2+i\omega\Gamma}\right)\:,\:\omega_p^2=\frac{e^2n}{\epsilon_b \epsilon_o m^*}
\end{equation}
with plasma wavelength $\lambda_p=2\pi c/\omega_p=\SI{5.7}{\um}$, scattering rate $\Gamma=1\times10^{13}\SI[per-mode=symbol]{}{\radian\per\second}$, and background permittivity $\epsilon_b=12.3$. The absorber region is modeled as having a constant real permittivity and an imaginary permittivity extracted assuming an absorption coefficient $\alpha(\omega)=\alpha_o\sqrt{E-E_g}$ where $E_g=\SI{248}{\milli\eV}$ is the bandgap of the MWIR T2SL and the factor $\alpha_0=\SI{7500}{\per\cm}$ scales the T2SL absorption.

We utilize Rigorous Coupled Wave Analysis (RCWA) \cite{moharam1995formulation} to optimize the device geometry (Fig. \ref{fig:1}(a)) including grating height ($h$), period ($\Lambda$), and grating width ($w$), and maximize absorption of the incoming light within the absorber layer of the detector. This optimization results in a geometry with $\Lambda=\SI{1.4}{\um}, h\simeq 400nm$, and grating ridge duty cycle $DC=\frac{w}{\Lambda}\times 100\%\simeq 36\%$. Fig. \ref{fig:1}(c) shows the enhancement in total absorption in the optimized GMR structure compared to its planar counterpart. 

To gain some insight into the observed behavior, we consider the detector to be a combination of planar waveguides with PEC and low-index n\textsuperscript{++} InAsSb claddings and different thicknesses of GaSb layers representing etched (ridge) and unetched (groove) parts of the grating. The dispersions of the guided modes in these waveguides,  analyzed with the transfer matrix method (TMM),  and  represented by the propagation constant ($k_{\rm GMR}$), are illustrated in Fig. \ref{fig:1}(b). The same figure illustrates the dispersion of the effective-medium mode whose propagation constant is a weighted-average of the propagation constants of the two guided modes mentioned above. Note that since $k_{\rm GMR}>2\pi/\lambda_0$, the guided modes cannot be directly excited by the free-space light. However, it becomes possible to couple normally-incident light to this mode using the diffraction grating (implemented into the PEC contact) when  $k_{\rm eff}\simeq 2\pi/\Lambda$. The main maximum in the absorption spectrum (Fig. \ref{fig:1}(c)) reflects such resonant coupling.

The additional absorption maxima (Fig. \ref{fig:1}(c)) can be understood as the Fabry-Perot-type resonances of normally incident light. The absorption inside the active layer of the device (our estimate for EQE) is shown in Fig. \ref{fig:1}(d) for a GMR detector (black) and its planar analog (brown). It is seen that GMR structure increases EQE by a factor of $\sim 30$. Note that the location of the peak in EQE coincides with the location of the peak in total absorption, while the difference between the spectral response of active layer and the n\textsuperscript{++} InAsSb cladding, along with the peculiarities of the intensity distribution across the stack, modify the contribution of the Fabry-Perot resonances to the performance of the final device. Overall, the ultra-thin (${\lambda_0}/{16}$) active layer is capable of absorbing approximately 60\%  of the incoming light. 

\subsection{Finite pixels}

 \begin{figure}[h!]
    \centering
    \includegraphics[width=\textwidth]{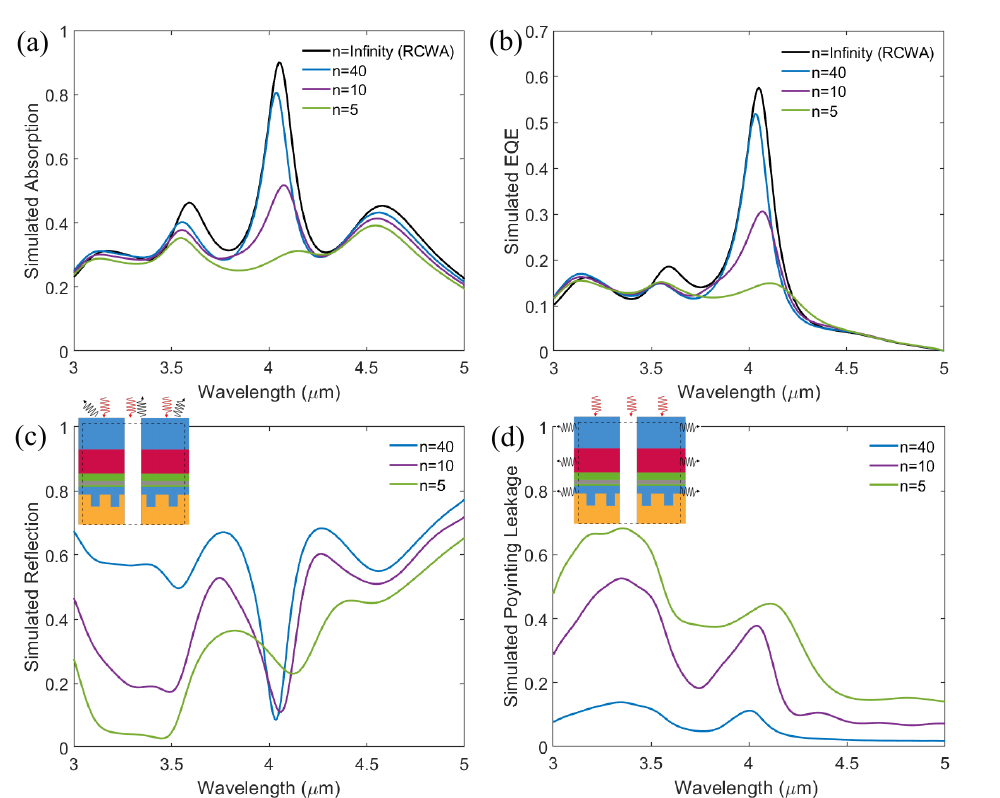}
    \caption{Simulated (a) total absorption and (b) external quantum efficiency (EQE) of the mid-wave IR GMR-based detector pixels a function of the number of grating periods ($n=\infty$, $n=40$, $n=10$, and $n=5$). Simulated (c) reflection and (d) Poynting flux leakage in finite-sized detectors; insets illustrate the directions of incident (red), and outgoing (black) energy fluxes in (c,d) }
    \label{fig:2}

\end{figure}

The calculations  above assume infinitely periodic detectors. Grating structures of actual practical devices, however, have a finite number of periods. The performance of such finite structures is analyzed with the commercial finite-element-method partial differential equation solver COMSOL Multiphysics \cite{comsol} that uses perfectly matched layers to mimic the open boundary conditions surrounding the detector pixel. As before, we use the absorption within the active region of the absorber layer (normalized to the power incident on this active region) as our estimate of the EQE of the final device. 

Figures \ref{fig:2}(a,b) show the simulated total absorption and EQE of finite MWIR GMR detectors with varying lateral size (characterized by the  number of periods in the grating structure $n$).Notably, the performance of the large-area detectors ($n=40$) is virtually identical to that of the infinite periodic structures, consistent with previous GMR demonstrations \cite{AKGMR1,AKGMR_2}. However, as the size of the detector decreases, its performance weakens significantly. For example, the performance of the $n=5$ detector (lateral width of $\approx\SI{7}{\um}$) approaches that of the planar T2SL stack, virtually eliminating any gains of the GMR architecture. 

The degradation of detector performance (EQE) with decreasing pixel size is further illustrated in Fig. \ref{fig:2}(c) which shows the percentage of light back-reflected from the detector active area. It is seen that as the number of grating periods decreases, the coupling to the guided mode becomes less efficient. However, this process is very inhomogeneous: the $n=10$ structure seems to couple incoming light into the GMR mode almost as efficiently as its larger ($n=40$) counterpart, while the coupling efficiency drops significantly between $n=10$ and $n=5$. To further understand the origin of the observed behavior, we calculated the energy flux that leaves the active area of the detector for a GMR mode. Such Poynting flux leakage is illustrated in Fig. \ref{fig:2}(d). The light coupled by the large ($n=40$) grating is either reflected or is almost completely absorbed within the device. As the grating count is reduced below the characteristic scale of GMR-mode propagation, a significant portion of the energy incident upon the detector leaves the device in the lateral direction. While this energy may be absorbed outside the active region of the device, such a configuration is not conducive to FPA operation. The electron-hole pairs excited by the leaking light will either not be collected or alternatively, will be collected by adjacent pixels (resulting in cross-talk), depending on the detector array geometry.  

\subsection{Cavity-Assisted Pixels}

\begin{figure}[t!]
    \centering
    \includegraphics[width=\textwidth]{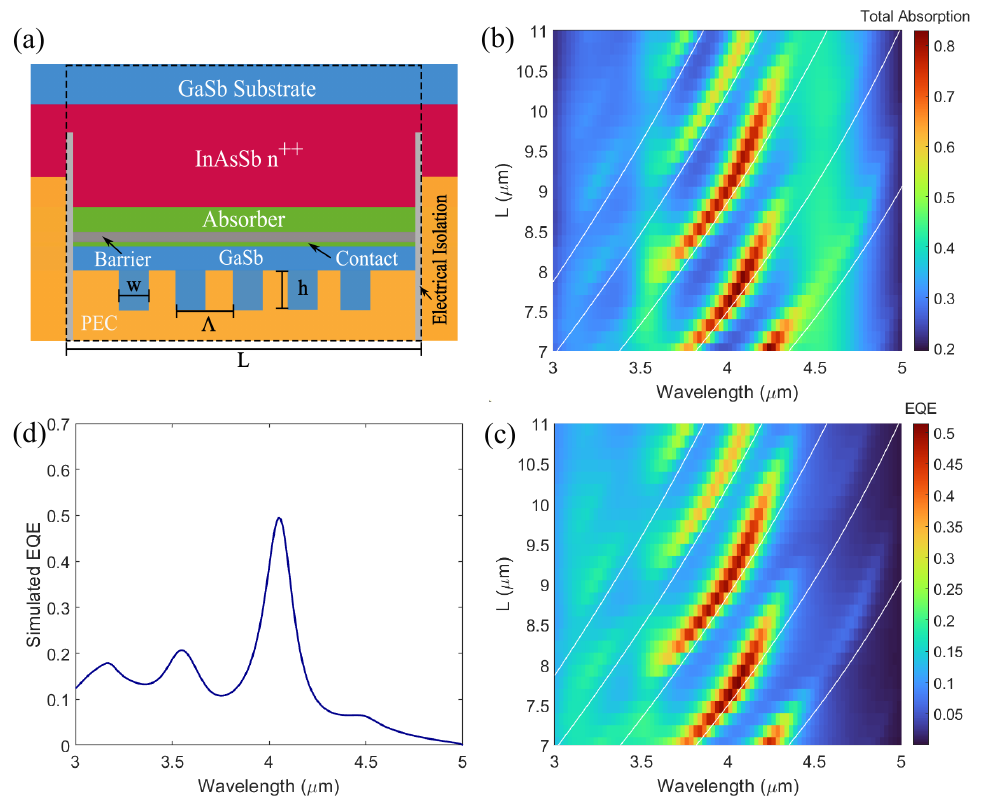}
    \caption{(a) Schematic design of the proposed resonant pixel with dashed box shows indicating the detector active region. Simulated (b) total absorption and (c) EQE of the device with $n=5$ as function of cavity width (L) and operating wavelength; dashed lines represent resonant excitation of GMR modes; (d) Spectral response of the device with $L=\SI{7.74}{\um}$ that relies on the combined GMR and cavity resonances.}
    \label{fig:3}

\end{figure}

In order to overcome the limitations of quasi-planar, finite-width detectors, we propose to introduce a lateral cavity structure designed to trap the light coupled into guided modes within the detector pixel. The design of the proposed structure that incorporates a second Au contact that extends to the $n^{++}$ ground plane, electrically insulated from the top contact, is shown in Fig. \ref{fig:3}(a).
In practice, fabrication of such a device would require depositing a thin ($\ll\lambda_0$) electrically-insulating layer that would have negligible effect on the overall optical response of the device, by either atomic layer deposition (ALD) or plasma-enhanced chemical vapor deposition (PECVD), prior to the second metal deposition step. 

The absorption within the structure as a function of wavelength and cavity size is shown in Fig. \ref{fig:3}(b), revealing a set of maxima that are controlled by the overall cavity size and wavelength. The thin lines in Fig. \ref{fig:3}(b) correspond to $L=(2m+1)\lambda_{\rm GMR}/2=\pi(2m+1)/k_{\rm eff}$ with integer $m$, illustrating that these absorption maxima result from the interplay between the lateral cavity modes and the guided modes of the structure. 

Similarly to what was shown for the infinite planar detectors above, the predicted EQE closely follows the overall absorption spectrum, with wavelength-dependent scaling that reflects different absorption within the active layer and the heavily doped ground plane. Fig. \ref{fig:3}(d) further illustrates  the performance of the $\approx\SI{8}{\um}$-wide cavity, demonstrating EQE $\approx51\%$. Notably, the performance of this compact structure is comparable to the much longer GMR detector, and is almost an order of magnitude larger than the performance of the compact quasi-planar GMR-only structure. 

\subsection{Alternative Pixel Designs}

\begin{figure}[t!]
    \centering
    \includegraphics[width=\textwidth]{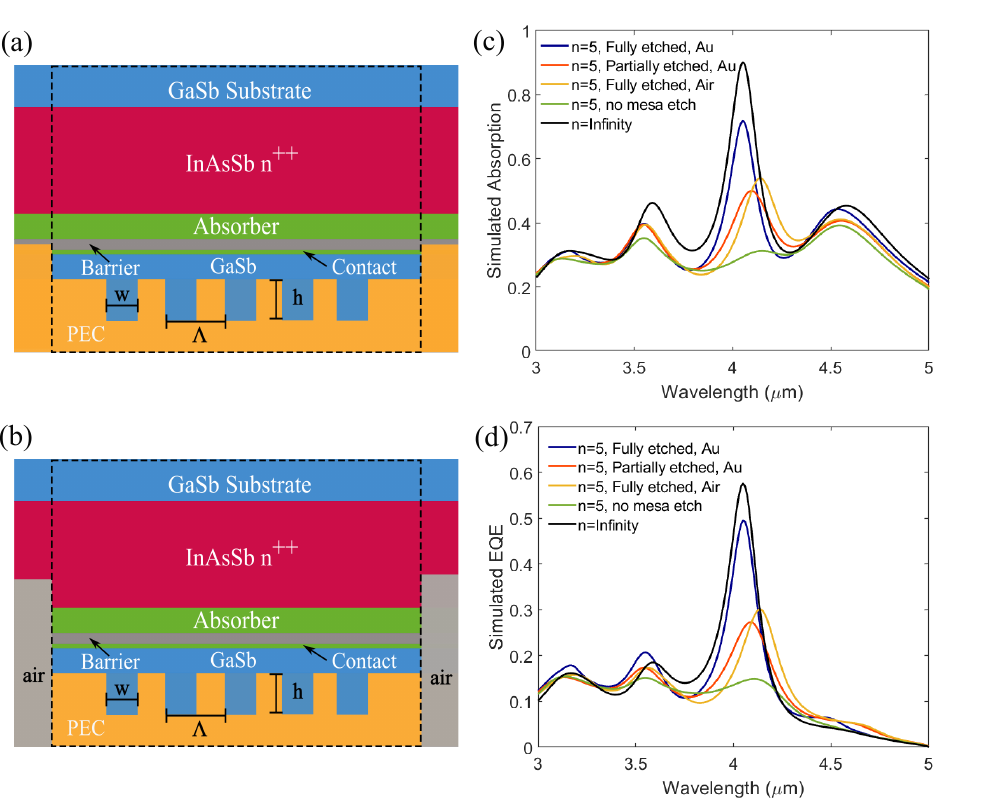}
    \caption{Simplified resonant pixel designs with (a) partial etch and (b) partial metallization; panels (c,d) illustrate the total absorption and EQE of simplified designs, respectively, and compare the performance of the simplified designs to that of fully-metalized resonant pixels, infinite GMR detectors, and planar unpatterned detectors}
    \label{fig:4}

\end{figure}
Fabricating the idealized pixel geometry, including vertical sidewalls and two highly conductive but electrically insulated contacts, may be challenging. We therefore consider performance of two alternative designs of the finite pixel that are easier to realize in experiments. These designs, schematically shown in Fig. \ref{fig:4} include (i) partial etching where the gold contact extends to the buffer layer (Fig. \ref{fig:4}a)  and (ii) full mesa etching with only the upper portion covered with gold (Fig. \ref{fig:4}b). 

In both designs, the metal layer provides one of the electrical contacts to the device; the second contact is to be fabricated in the far field of the active region by electrically contacting the n\textsuperscript{++} plane. However, the convenience of these designs comes at the cost of reduced quality factors of the cavity resonance, as compared with the design shown in Sec.2.3.  

The performance of the two alternative designs is illustrated in Fig. \ref{fig:4}(c-d). As expected, the absorption (Fig. \ref{fig:4}), as well as the predicted EQE, of the alternative setups are in-between those of quasi-planar and fully-metalized devices. From practical perspective, the 30\% EQE, predicted in both alternative designs represents a significant improvement over the quasi-planar structure. We expect that larger (10-period) structures may exhibit a reduced trade-off between idealized and alternative designs. 

\section{Experiment and Analysis}
\subsection{Fabrication and Characterization}
An experimental set of lateral cavity-enhanced GMR pixels were grown by molecular beam epitaxy as detailed in Section 2.1. After growth, these detectors were fabricated into $\SI{300}{\micro\meter}\times\SI{300}{\micro\meter}$ patterns of repeated 1D pixels, with a $\approx\SI{4}{\micro\meter}$ spacing between neighboring mesas. The gratings were patterned with $\Lambda\approx\SI{1.4}{\micro\meter}$ using electron-beam lithography and were inductively-coupled plasma etched. Pixel mesas were then fabricated with a second electron-beam lithography and etch, with patterns that vary the size and lateral position of the mesas with respect to the gratings. 
For fully-optical pixels, we etch through the core of the waveguide and into the ``$n^{++}$'' which results in a stronger lateral cavity. An atomic-layer deposition step was then performed to coat the samples in a conformal $~\SI{10}{\nano\meter}$ of alumina. This step ensures that the deposited metal could not react with any layers in the epitaxial stack. Finally, the arrays of pixels are coated with \SI{800}{\nano\meter} of silver to ensure that the previous etches would be back-filled and that minimal light would leak from the fully-etched pixels. 

\begin{figure}[t]
    \centering
    \includegraphics[width=\textwidth]{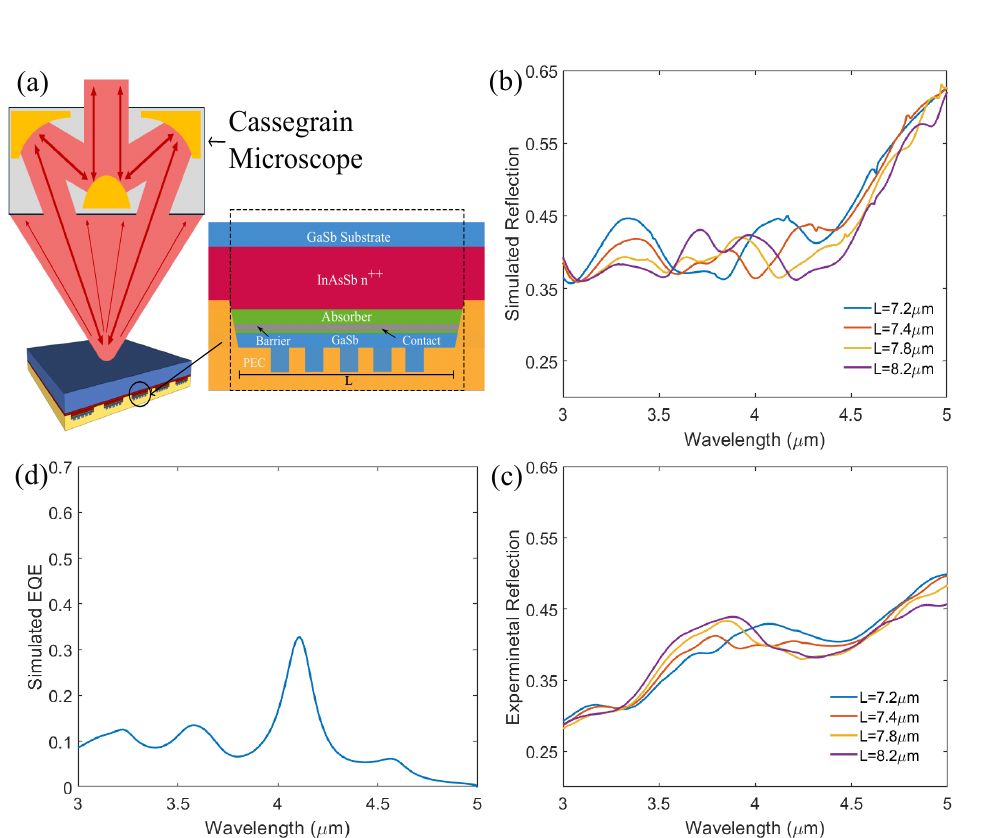}
    \caption{(a)  Schematic of the optical characterization setup and geometry of the finite pixel realized in experiments. (b) Simulated total reflection of periodic arrays of pixels with varying cavity width (c) Experimentally acquired reflection; (d) Simulated EQE of as-fabricated pixel under normal-incident illumination  with anti-reflection coating at GaSb-Air interface}
    \label{fig:5}

\end{figure}

After fabrication, the reflection of the detectors is measured using a Bruker V80 Fourier-transform infrared (FTIR) spectrometer, coupled to a Bruker Hyperion-II microscope with a motorized xy-stage and a MCT photodetector (Fig. \ref{fig:5}(a)). The setup is configured to measure reflection and is manually adjusted such that the microscope's focal spot overlaps with the pixel elements through the GaSb substrate. Great care was taken to ensure the focal spot overlapped with the arrays of pixels, as alignment through the substrate cannot be done visually. The resulting reflection spectra for detectors of several lateral width are shown in Fig. \ref{fig:5}(c). 

\subsection{Theoretical modeling and Discussion}

Note that the measurement and analysis are complicated by the Cassegrain reflector used as the microscope objective (Fig. \ref{fig:5}(a)). Overall, the experimental setup (aimed at optical characterization of compact pixels) has several intrinsic limitations: (i) the device is illuminated obliquely  (light is incident on the GaSb substrate from air at an angle between $7^{\circ}$ and $11^{\circ}$), (ii) illumination often encompasses multiple periodically arranged pixels, and (iii) overall reflection is strongly affected by the reflection from the substrate-air interface. (As mentioned above, we envision that practical detectors will incorporate anti-reflection coatings and focusing elements directing normally-incident light towards compact pixels). Finite elements modeling is used to incorporate the experimental conditions into our numerical analysis. As compared with design described above, we incorporate periodic boundary conditions and modify the direction of the incoming beam in our FEM model. Finally, we utilize intensity-based transfer matrix to estimate the reflection from the air-(optically-thick-)substrate interface. 

The reflection spectra, obtained with this procedure are shown in Fig. \ref{fig:5}(b). It is seen that theoretical simulations are in agreement with experimental data. Indeed,  overall reflection magnitude is in the same range, with each pixel exhibiting multiple reflection minima. The minima at $\lambda_0\simeq 4\mu m$ represent coupling to the hybrid GMR-cavity resonances. Apart from the overall shift of the order of $\sim 0.2 \mu m$ between experiments and theory, the spectral dependence of the  main absorption peak on the cavity size is consistent between experimental results and theoretical predictions. We attribute the above spectral shift to  deviations of real dimensions of experimental samples from their nominal values used in theoretical simulations. 

Having obtained reasonable agreement with the experimental data, we used the procedure described above to simulate the predicted performance of an individual pixel with nominal experimental dimensions, by assuming perfect anti-reflection setup at the air/substrate interface, and normalizing the absorption to the energy flux that directly enters the pixel. The results of this analysis are shown in Fig. \ref{fig:5}(d), demonstrating EQE of $\sim33\%$ in as-fabricated structures.

\section{Conclusion}
To conclude, we have presented designs of truly compact mid-wave-IR photodetector pixels that take advantage of hybridized guided mode and cavity resonances, predicting EQE of $\sim50 \%$ in 2-wavelength-wide, wavelength/15-thick structure. The proposed designs have been implemented in monolithically grown epitaxial material platform. Results of optical characterization are consistent with theoretical predictions. 

\begin{acknowledgments}
The authors gratefully acknowledge funding from the NASA Advanced Component Technologies program, Award No. ACT-22-0041. This work was partially conducted at the Texas Nanofabrication Facility supported by the NSF under Grant No. NNCI-202522.
\end{acknowledgments}

\bibliography{ResGMRrefs}

\end{document}